\documentclass[12pt,preprint]{aastex}

\usepackage{graphicx}
\usepackage{color}
\usepackage{amsmath}
\usepackage{natbib}

\shorttitle{The IR Afterglow of SMBH Mergers}
\shortauthors{Schnittman and Krolik}

\begin{document}

\title{The Infrared Afterglow of Supermassive Black Hole Mergers}
\author{Jeremy D.\ Schnittman\altaffilmark{1} and Julian H.\
  Krolik\altaffilmark{2}}
\altaffiltext{2}{Johns Hopkins University; {\tt schnittm@pha.jhu.edu}}
\altaffiltext{2}{Johns Hopkins University; {\tt jhk@pha.jhu.edu}}

\begin{abstract}
We model the spectra and light curves of circumbinary accretion disks
during the time after the central black holes merge. The most
immediate effect of this merger is the deposition of energy in the
disk due to the gravitational wave energy and linear
momentum flux released at merger. This has the effect of perturbing
the circular orbits of gas in the disk, which then intersect and
radiate the dissipated energy.
Because the disk is expected to be very optically
thick, the radiation emerges predominantly in the infrared, and
lasts for tens of thousands of years when the total black hole mass is
$M \sim 10^8 M_\odot$. On the basis of a simple cosmological merger model
in which a typical supermassive black hole undergoes a few major mergers
during its lifetime, we predict that $\sim 10^4-10^5$ of these afterglow
sources should be observable today.  We also discuss the possibility of
identifying them with existing multi-wavelength surveys such as
SWIRE/{\it XMM}-LSS/XBootes and COSMOS.
\end{abstract}

\keywords{black hole physics -- galaxies: nuclei}

\maketitle

\section{INTRODUCTION}
\label{intro}

Recent results from numerical relativity suggest that the merger of
two rapidly spinning black holes (BHs) can result in a large recoil from the
anisotropic emission of gravitational waves
\citep{bekenstein73,fitchett83,baker06,baker07,campa07a,campa07b,gonza06,
gonza07,herrm06,herrm07,tichy07}. This recoil, or ``kick,''
may lead to a number of interesting astronomically observable
signatures, many of them much longer lived than the gravitational wave
signal itself.
Some of these signatures are indirect, and depend on
galactic dynamics effects involving the merged black hole.
For example, low-density cores in the central regions of galaxies may indicate
a kicked black hole that ejected stars as it relaxed back to the
center via dynamical friction
\citep{merri04,gualandris07}. Alternatively, if the merged BH is ejected from the galaxy, it
may also leave behind a bulge without changing its mass or velocity
dispersion, thus breaking the $M$-$\sigma$ relation for a number of systems
\citep{volonteri07}. The overall population fraction of
central black holes may give a way of indirectly measuring the
distribution of BH recoil velocities \citep{schnittman07b}.

Another class of recoil observations depends on the direct
electromagnetic (EM) signature from gas accreting onto the kicked BH.
The quasar that results may be spatially displaced
from the galactic center \citep{loeb07},
and its line spectrum may be Doppler shifted relative to the host
galaxy \citep{bonning07}. \citet{milos05} discuss the EM signal
that would appear following a BH merger within a circumbinary gas
disk, in which a central gap of gas was cleared out by the inspiraling
BHs, but then is refilled on an inflow time after the merger. This
scenario, independent of any recoil kick, could provide a soft X-ray
counterpart to a LISA event on
a timescale of a few months to years after the gravitational wave (GW) signal.
More recently, \citet{lippai08} investigated the emission from shocks in a
disrupted disk in a similar configuration to that of \citet{milos05},
but including a kick for the final BH.  \citet{kocsis08} estimated
the (small) luminosity generated by viscous dissipation associated with
the passage of the gravitational wave pulse itself through a surrounding
accretion disk.  Within this class of direct EM
emission models, new speculation predicts the possibility of trails of gamma-ray
emission from dark matter annihilation in the wake of the recoiling BH
\citep{mohayaee08}.

In this paper we also consider a circumbinary disk around an
inspiraling binary BH system, but focus on the long-term
afterglow emission from the perturbed disk. At the time of merger, the
gas in the surrounding disk is instantaneously placed on
eccentric orbits due both to the change in gravitational potential from GW
energy loss and the recoil of the central BH. These perturbed
orbits relax back to circular trajectories while conserving
angular momentum but dissipating energy over a few orbital
periods. Because these disks are very optically thick, the dissipated
energy will be radiated with a thermal spectrum, giving
a warm yet relatively short-lived signal from the inner edge of the disk,
followed by prolonged infrared (IR) emission from the outer regions
of the disk. This signature differs from that predicted by \citet{lippai08}
and \citet{shields08}, who argued that there would be prompt
emission in the ultraviolet (UV) and X-ray bands, based on
calculations that do not include the disk opacity.
Our work also contrasts with previous efforts in that both
\citet{milos05} and \citet{lippai08} focused on EM counterparts to LISA
sources, and thus limited the total mass in the merger to $\lesssim 10^6
M_\odot$ \citep{kocsis07}. We relax this limit
because observations suggest that the SMBH mass function
$dN/d \log M$ peaks closer to $\sim 10^8 M_\odot$ for
redshifts $z\gtrsim 0.5$ \citep{marconi04,merloni04}.
However, we also consider the lower mass range so that
we can estimate the character of the afterglows produced by the
smaller mass mergers that may produce gravitational
waves detectable by LISA.

For black hole masses of $\sim 10^8 M_\odot$, the IR afterglow
signature could last for hundreds of
thousands of years and provide evidence of SMBH mergers even
without LISA signals. For reasonable pre-merger accretion rates,
the central gap cleared out by the inspiraling binary does
not close for $\sim 10^6$ years, giving a unique signature
with infrared luminosity comparable to that of a quasar, and virtually
no UV or X-ray emission. Similarly, unlike
classical AGN, we expect no compact radio jets or photoionized emission lines.
Using the anti-hierarchical BH mass function of \citet{merloni04} and
an estimate of the SMBH merger rate \citep{sesana04}, we
expect a total merger rate of $\sim 0.1$ per year out to $z\sim
6$ for total mass $M > 10^6 M_\odot$. Due to the long
lifetimes of such afterglows, as many as
$\sim 10^4-10^5$ may be observable in the entire sky at any one
time. Within the fields and flux sensitivities
of existing multi-wavelength surveys such as SWIRE and COSMOS, we expect
that $\sim 1-10$ may be discernable today, but any candidates would
require follow-up optical spectroscopy for positive identification.

This paper is organized as follows: in Section \ref{merger_model} we
describe a simple model for a binary merger including GW
energy and momentum losses and the subsequent behavior of the
perturbed circumbinary disk, with predictions of light curves and
spectra. In Section \ref{cosmological_model} we combine these results
with a cosmological merger model and calculate expected merger rates
and distribution of afterglow spectra. In Section
\ref{observations} we discuss the feasibility of observing such
systems, and present our conclusions in Section \ref{discussion}.

\section{SPECTRAL SIGNATURE OF A SINGLE MERGER}
\label{merger_model}

\subsection{Circumbinary disk model}\label{disk_model}

We begin with a circumbinary disk described by an inner radius
$R_{\rm in}$, total BH mass $M = M_1+M_2$, mass ratio $q=M_1/M_2\le 1$,
accretion rate $\dot{M}$, and an efficiency
$\eta$ such that, if the disk extended all the way in to a single
central black hole, the luminosity would be $L=\eta \dot{M} c^2$. For
this efficiency, we can define a normalized accretion rate
$\dot{m}=\dot{M}/\dot{M}_{\rm Edd}$, where $L_{\rm Edd}=\eta
\dot{M}_{\rm Edd}c^2$ is the Eddington luminosity. Outside of $R_{\rm in}$, we
describe the disk by a steady-state alpha model, with nominal
efficiency $\eta=0.1$ and viscosity parameter $\alpha=0.1$
\citep{shakura73}. The sensitivity of our results to these assumptions
is discussed in Section \ref{discussion}.

For a pair of BHs on a circular orbit with binary separation $a$, the
inner edge of the circumbinary disk is located at
  $R_{\rm in}=\lambda a$, where $\lambda(q)$
is a function only of the mass ratio and exhibits a relatively small
range around $\lambda \approx 1.6-1.8$ for $q\gtrsim 0.01$
\citep{artymowicz94}. Interior to this point, gas either
accretes directly onto one of the two BHs or, more likely, is ejected
from the system, giving a gap of very low density inside of $R_{\rm
  in}$ \citep{macfadyen08,bogdanovic08}. When the
binary separation is sufficiently small, its evolution becomes
dominated by gravitational wave losses and inspirals on a timescale
given by the leading-order quadrupole formula \citep{peters64}:
\begin{equation}\label{t_inspiral}
t_{\rm insp}(a) = -\frac{a}{\dot{a}} = \frac{5}{64}\frac{c^5}{G^3}
\frac{a^4}{M^2\mu},
\end{equation}
where $\mu=M_1 M_2/M$ is the reduced mass of the binary.

As the binary orbit shrinks, the inner edge of the disk follows
closely behind, maintaining the relation $R_{\rm in}=\lambda a$ as
long as the inflow time is less than the inspiral time.
For a standard alpha disk, the gas inflow time is given by
[e.g.\ \citet{krolik99}]
\begin{equation}\label{t_inflow1}
t_{\rm inflow}(R) \approx \frac{R^2\Omega_{\rm orb}}{\alpha c_s^2}
\approx \frac{1}{\alpha} \frac{R^{3/2}}{(GM)^{1/2}}
\left(\frac{R}{h}\right)^2,
\end{equation}
where $\Omega_{\rm orb}$ is the orbital angular velocity, $c_s$ is the
sound speed, and $h$ is the disk thickness. In the inner regions of
the disk, where the pressure is radiation-dominated and the opacity is
dominated by electron scattering, we have
\citep{shakura73, novikov73}
\begin{equation}\label{t_inflow2}
t_{\rm inflow}(R_{\rm rad}) \approx 5 \times 10^6\, \alpha_{-1}^{-1}
\eta_{-1}^2 \dot{m}_{-1}^{-2} M_8 x_3^{7/2} \mbox{ yr},
\end{equation}
while in the outer, gas-pressure and scattering-dominated region,
we have 
\begin{equation}\label{t_inflow3}
t_{\rm inflow}(R_{\rm gas}) \approx 7 \times 10^5\, \alpha_{-1}^{-4/5}
\eta_{-1}^{2/5} \dot{m}_{-1}^{-2/5} M_8^{6/5} x_3^{7/5} \mbox{ yr}.
\end{equation}
Here we have employed scaled parameters $\alpha=0.1\alpha_{-1}$,
$\eta=0.1\eta_{-1}$, $\dot{m}=0.1\dot{m}_{-1}$, $M=10^8\, M_8\,
M_\odot$, and $x=10^3 x_3(Rc^2/GM)$ is the radius in geometric
units. The point where the disk transitions from radiation- to
gas-pressure dominated can be estimated by setting $t_{\rm
  inflow}(R_{\rm rad})=t_{\rm inflow}(R_{\rm gas})$, or
\begin{equation}
x_{\rm trans} \approx 4\times 10^2\, \alpha_{-1}^{10/105}
\eta_{-1}^{-4/5} \dot{m}_{-1}^{4/5} M_8^{10/105}.
\end{equation}
Scaled in terms of $x_3$, the inspiral time is
\begin{equation}\label{t_inspiral2}
t_{\rm insp}(R_{\rm in}/\lambda) \approx 1 \times 10^6\, M_8\,
\frac{(1+q)^2}{q}\lambda^{-4}(q) x_3^4 \mbox{ yr}.
\end{equation}

When the inspiral time becomes shorter
than the gas inflow time at the inner edge,
the binary separation decreases faster than the gas can move in,
effectively decoupling the BHs from the disk, and the
merger occurs soon after \citep{milos05}.  For our nominal
model parameters $M=10^8 M_\odot$,
$q=1$, $\eta=0.1$, and $\dot{m}=0.1$, the central gap in the disk is
quite large: $R_{\rm in}\approx 10^3M$. However, we also
find that for some system parameters, namely when $q$
is small (thus a long inspiral time) and the accretion rate is large
(thus a short inflow time), the inner edge of the disk is able to keep
up with the gravitational inspiral up to the final plunge and no
appreciable gap is formed.  Furthermore, we find that whenever a gap
{\it is} formed, the inner edge is almost always within the
gas-pressure dominated region of the disk, so we can solve for
$R_{\rm in}$ with equations (\ref{t_inflow3}, \ref{t_inspiral2}):
\begin{equation}
(R_{\rm in}/M) \approx 1 \times 10^3\, \alpha_{-1}^{-4/13}
\eta_{-1}^{2/13} \dot{m}_{-1}^{-2/13} M_8^{1/13}
\left[\frac{q\lambda^4(q)}{(1+q)^2}\right]^{5/13}.
\end{equation}
(Note that in \citet{milos05}, the inflow time is
taken to scale as $R^2$, and they consequently find smaller values of
$R_{\rm in}$, typically within the radiation-dominated region of the disk.)

\subsection{Post-merger dynamics of a perturbed disk}

At the time of merger, $M$ is reduced nearly instantaneously
by a fraction $\epsilon_{\rm GW}$ due
to GW energy flux, with $\epsilon_{\rm GW}$ typically a few percent. This mass loss
is scaled to the symmetric mass ratio $\nu \equiv q/(1+q)^2$,
normalized to numerical simulations of non-spinning BH mergers
\citep{gonza06,herrm06}:
\begin{equation}\label{e_GW}
\epsilon_{\rm GW} \approx 0.08\, \nu +0.32\, \nu^2.
\end{equation}
This mass loss could be as high as $\epsilon_{\rm GW}\approx 0.1$ for
aligned, rapidly spinning BHs \citep{marronetti07,dain08}, but averaging over
an ensemble of uniformly distributed spin orientations, we think the
non-spinning estimate is reasonable.

For generic systems without special symmetries, the resulting black hole also
receives a linear momentum recoil with velocity $V_{\rm kick}$.
This recoil leaves the outer regions of the
disk (where $V_{\rm orb}\lesssim V_{\rm kick}$) unbound and significantly
disrupts the inner regions. For the bound regions ($V_{\rm orb}\gtrsim
V_{\rm kick}$), an
annulus of mass $dM(R)=2\pi R\, dR\, \Sigma(R)$ on an originally circular
orbit at radius $R$ receives an instantaneous boost in energy
relative to the merged black hole of
\begin{equation}\label{dE}
dE_{\rm kick}(R_{\rm bound})=\frac{G\, dM(R)\, M}
{R} \epsilon_{\rm GW} +\frac{1}{2}dM(R)\, V_{\rm kick}^2.
\end{equation}
In the unbound outer regions, the change in energy is simply the
original binding energy $dE_0$:
\begin{equation}\label{dE_2}
dE_{\rm kick}(R_{\rm unbound}) = -dE_0=\frac{G\, dM(R)\, M}{2R}
=\frac{1}{2}dM(R)\, V_{\rm orb}^2.
\end{equation}

In the inner-most regions, where the mass-loss contribution to the energy
gain is significant, it is possible that not all this added energy
will be available for dissipation.  This is because (as pointed out
to us by Cole Miller), in the perfect (i.e. dissipationless) fluid
limit, fluid elements retain their initial angular momentum.  With
the new, smaller central mass, that angular momentum is too large
for any bound orbit at their initial radius.  If fluid dynamics
enforces circular orbits in the long-run, individual fluid elements
expend most of the energy they gained by the mass-loss on moving
out to their new orbital radius, leaving only $\sim (GM/R)\epsilon_{GW}^2$
available for dissipation.  However, the actual mechanics of this
response may well involve numerous shocks in which considerable
angular momentum mixing between different fluid streams can occur.
The effectiveness of this mixing will depend on numerous considerations
including the radial surface mass density profile, the rapidity of
cooling, and the role of radiation pressure (see below for further
estimates relevant to the latter two), and its evaluation is therefore
best left to future work.  Given this uncertainty, we have opted for
the more optimistic scaling (heating from mass-loss $\sim \epsilon_{GW}$
times the binding energy) in the results shown here.  However, we have
checked that whether one adopts a mass-loss heating scaling $\sim \epsilon_{GW}$
or $\sim \epsilon_{GW}^2$ makes very little difference to our
conclusions because it alters the afterglow at an
interesting level only during the first $10^2$--$10^3$~yr after merger,
while nearly all the detectable objects will be seen at much later times.

For the systems we have considered, the vast majority of
the afterglow energy is released in the regions of the disk that are dominated
by gas pressure (typically $R\gtrsim 10^3M$), where the surface density
is given by \citep{shakura73,novikov73}
\begin{equation}\label{Sig_r}
\Sigma(R) \approx 1\times 10^5\alpha_{-1}^{-4/5}
\eta_{-1}^{-3/5} \dot{m}_{-1}^{3/5} M_8^{1/5} x_4^{-3/5}\mbox{ gm/cm$^2$}
\end{equation}
and the gas density is
\begin{equation}\label{rho_r}
\rho(R) \approx 1\times 10^{-10}\alpha_{-1}^{-7/10}
\eta_{-1}^{-2/5} \dot{m}_{-1}^{2/5} M_8^{-7/10} x_4^{-33/20}\mbox{ gm/cm$^3$}.
\end{equation}
From here on, we use a fiducial radial distance of $10^4$
gravitational radii, corresponding to the marginally bound region of the
disk for the largest expected kicks.
Strictly speaking, equations (\ref{Sig_r}, \ref{rho_r}) assume that
the opacity is dominated by electron scattering, 
but they are actually quite similar to the expressions corresponding
to free-free/bound-free scattering. In practice, both opacity effects
tend to be important (and also dust and molecular opacities at lower
temperatures in the outer-most regions of the disk)
and we find our pre-merger disks are extremely optically
thick out to radii of at least
$x=10^5$. Since the total energy released at each point in the disk
is proportional to the surface density (eqn. \ref{dE}), only optically
thick disks will actually have enough mass to produce a significant
amount of luminosity. Moreover, because $\Sigma(R)$ depends on
$\dot{m}$ only to the $3/5$ power and on $M$ only to the $1/5$ power,
our conclusion that these disks have large optical depths should be
valid over a wide range of possible conditions. As we
will show below, even the abrupt heating that takes place after the black
hole merger does not alter this conclusion.

The change in energy given in equation (\ref{dE})
leads to a perturbed disk with
eccentric, intersecting orbits. Conserving angular
momentum in each annulus, the gas relaxes back to a collection of
circular orbits on a timescale of order $t_{\rm heat}$ as shocks due
to intersecting orbits dissipate energy in the disk. We define
the heating time as the time it takes for nearby eccentric orbits to
cross. In the epicyclic approximation for small eccentricity $e$, the
radial coordinate $R$ of a perturbed orbit is
\begin{equation}
R(t)=R_0(1+e\cos\psi),
\end{equation}
where $R_0$ is the radius of the guiding center orbit and
$\psi=\Omega_{\rm orb}t+\psi_0$ is the epicyclic phase. Density caustics form when orbits
with different $R_0$ overlap at the same $R$:
\begin{equation}
\frac{\partial R}{\partial R_0}
= \frac{3}{2}\frac{(GM)^{1/2}t}{R_0^{3/2}} (e\sin\psi)+(1+e\cos\psi) = 0.
\end{equation}
In the limit of small $e$, we get a heating time of
\begin{equation}\label{t_heat1}
t_{\rm heat} \approx \frac{2}{3} \frac{R_0^{3/2}}{(GM)^{1/2}e}
 = \frac{1}{3\pi}\frac{t_{\rm orb}}{e}.
\end{equation}

In the simplest case where the kick velocity is zero,
the instantaneous GW mass-loss will excite the entire disk with a constant
eccentricity $e_{\rm mass-loss}\approx \epsilon_{\rm GW}$, independent
of radius. In the opposite regime where the mass-loss contribution to
the eccentricity is much smaller than that of the recoil, the post-kick
magnitude of the total specific angular momentum $\ell$ of a single
fluid element relative to the merged black hole will be
\begin{equation}\label{l_fin}
\ell^2 = R_0^2(V_{\rm orb}^2+2\beta V_{\rm orb}V_{\rm kick} + \beta^2
V_{\rm kick}^2+ z^2 V_{\rm kick}^2),
\end{equation}
where $\beta V_{\rm kick}$ is the component of the kick parallel to
the fluid velocity vector, $z V_{\rm kick}$ is the component of the kick
perpendicular to the orbital plane.
The semi-major axis of such a fluid element can be determined from its
specific energy $\varepsilon$:
\begin{equation}\label{a_fin}
\frac{1}{a} = -\frac{2\varepsilon}{GM}=\frac{1}{GM}(V_{\rm orb}^2-2\beta V_{\rm orb} V_{\rm kick}-V_{\rm
  kick}^2).
\end{equation}
Combining equations (\ref{l_fin},\ref{a_fin}), we find the perturbed
fluid element will have a post-kick eccentricity of
\begin{equation}\label{e_kick}
e_{\rm kick} = \left(1-\frac{\ell^2}{GMa}\right)^{1/2}
=\frac{V_{\rm kick}}{V_{\rm orb}} \left[(1+3\beta^2-z^2)
+\frac{V_{\rm kick}}{V_{\rm orb}}2\beta (1+\beta^2+z^2)
+\frac{V_{\rm kick}^2}{V_{\rm orb}^2}(\beta^2+z^2)\right]^{1/2}.
\end{equation}
This is an exact result for bound, Keplerian orbits. For $V_{\rm kick}
\ll V_{\rm orb}$, we find the typical eccentricity scales like $e
\approx V_{\rm kick}/V_{\rm orb}$ for planar kicks with $z=0$ and $e
\approx (V_{\rm kick}/V_{\rm orb})^2$ for kicks directed out of the
orbital plane with $z=1$ and $\beta=0$. While the final kick direction
is not independent of its magnitude, the relationship
is not well-known at this point \citep{tichy07}, so for these simple
estimates, we typically average over angles to get $\beta^2=z^2=1/3$.

In practice, when estimating
the heating rate in equation (\ref{t_heat1}), we use a combination of
the mass-loss and kick eccentricities:
\begin{equation}
e_{\rm tot} = \sqrt{\epsilon_{\rm GW}^2+\langle e_{\rm
kick}^2\rangle},
\end{equation}
where $\langle e_{\rm kick}^2\rangle$ is the average of equation
(\ref{e_kick}) over an isotropic distribution of kick directions. We
further impose the constraint that $e_{\rm tot} \le 1$ everywhere in
the disk, with $e_{\rm tot}=1$ indicating the unbound regions.

To calculate the light curves produced by the
merger, we must first estimate the temperature, density, and optical
depth of the perturbed disk.
During the heating phase, shocks with characteristic speed $V_{\rm shock}$
will be driven through the gas at each annulus in the disk.  In much
of the bound region, the shock speed will be roughly
limited by the kick speed, a result confirmed by \citet{shields08}, who
calculated the relative velocity of colliding geodesic particles on
perturbed orbits, finding $V_{\rm shock}\approx 0.3 V_{\rm kick}$ at
early times and $V_{\rm shock}\approx 0.9V_{\rm kick}$ at later
times. In the innermost regions of the disk,
the GW mass loss (neglected by \citet{shields08}) will dominate the
orbital dynamics, giving 
\begin{equation}
V_{\rm shock, bound} \lesssim
  \max\left[(2\epsilon_{\rm GW})^{1/2} V_{\rm orb},V_{\rm kick}\right].
\end{equation}
In the unbound regions, the shock speed is limited
by the orbital velocity, so 
\begin{equation}
V_{\rm shock, unbound} \approx
  \min\left[V_{\rm orb},V_{\rm kick}\right]
\end{equation}
everywhere in those regions. It should be noted that even if
the shock speeds in the bound regions are much smaller than the kick 
speed, as suggested by \citet{lippai08}, the total kick energy must
still be dissipated eventually, and only the initial temperatures will be
different. However, as we will see below, the post-shock disk should
be quite optically thick to thermalization, giving robust spectral
and light curve predictions independent of the specific heating
details.

\subsection{Dissipation of energy in the disk}

Assuming the pre-merger disk is relatively cold, the
initial shocks heat the gas to a temperature
\begin{eqnarray}
T_{\rm shock} &=& \frac{3}{16} \frac{\mu}{k} V_{\rm shock}^2\mbox{ K}
\nonumber\\
&\approx& 1.4 \times 10^7\, V_{1000}^2\mbox{ K},
\end{eqnarray}
where $\mu \approx 0.6 m_p$ is the mean molecular mass of the gas and
$V_{1000}$ is the shock velocity in thousands of km/s {\citep{mckee80}}.
This energy in turn will be transformed to radiation on the bremsstrahlung
radiation timescale:
\begin{equation}
t_{\rm rad}
= \frac{3kT_{\rm shock}}{1.4\times 10^{-27} (\rho/m_p) T_{\rm shock}^{1/2}}
\approx 17\, V_{1000} \, \alpha_{-1}^{7/10}
\eta_{-1}^{2/5} \dot{m}_{-1}^{-2/5} M_8^{7/10} x_4^{33/20} \mbox{ s},
\end{equation}
where we have used the free-free volume emissivity given in \citet{shapiro83}.
This timescale is much shorter
than any dynamical time in the system, so we assume all the
shock energy is converted instantly into radiation.

At the high temperatures prevailing immediately post-shock,
the disk may no longer be optically thick to
free-free absorption, but will still be highly opaque to electron
scattering.  Where, as here, $\tau_{es} > \tau_{ff}$, the optical depth
to thermalization is given by
$\tau_{\rm therm} = (\tau_{\rm es} \tau_{\rm ff})^{1/2}$, where $\tau_{\rm es}$ and
$\tau_{\rm ff}$ are the electron scattering and free-free optical
depths, respectively:
\begin{subequations}\label{tau_therm}
\begin{eqnarray}
\tau_{\rm es} &=& 0.35\, \Sigma, \\
\tau_{\rm ff} &=& \bar{\kappa}_{\rm ff} \Sigma \approx 2.3\times
10^{24}\, \rho T^{-7/2} \Sigma, \\
\tau_{\rm therm} &\approx& 0.5\, \alpha_{-1}^{-23/20} \eta_{-1}^{-4/5}
\dot{m}_{-1}^{4/5} M_8^{-3/20} x_4^{-57/40} T_7^{-7/4}.
\end{eqnarray}
\end{subequations}
Here $\bar{\kappa}_{\rm ff}$ is
the mean free-free opacity calculated by averaging over a
bremsstrahlung emission spectrum at temperature $T = 10^7\, T_7$ K.
Thus, for typical parameters, we expect the disk
immediately post-shock to be marginally optically thick in terms of
thermalization.

When the parameters are such that the initial
burst of free-free radiation is well-thermalized, the spectrum
quickly evolves through absorption and reradiation to Planck
form.  As it does so,
the matter and the radiation reach a state of thermodynamic
equilibrium at a temperature defined by matching the thermal
energy density created by the shock to the total heat content
of gas and photons:
\begin{equation}
\frac{9}{32}\rho V_{\rm shock}^2 = aT_{\rm therm}^4+\frac{3}{2}
\frac{\rho}{\mu} kT_{\rm therm}.
\end{equation}
For any shocks with $V_{\rm shock} \gtrsim 100$ km/s, the thermodynamic equilibrium
pressure will be completely dominated by the radiation.  The
equilibrium temperature is then
\begin{equation}\label{t_therm}
T_{\rm therm} \approx 8\times 10^4\,  V_{1000}^{1/2} \alpha_{-1}^{-7/40}
\eta_{-1}^{-1/10} \dot{m}_{-1}^{1/10} M_8^{-7/40} x_4^{-33/80}\mbox{ K}.
\end{equation}

The fact that the thermodynamic equilibrium is strongly dominated by
radiation pressure has the consequence that only a small portion of the
initially-radiated free-free emission need be immediately thermalized
in order for the gas quickly to reach thermodynamic equilibrium.
Because free-free opacity at frequencies $\nu$ well below $kT/h$ is
$\propto \nu^{-2}$, even when the blackbody peak is at first not thermalized,
sufficiently low frequencies are.  Thus,
for those regions with $\tau_{\rm therm}<1$, we can define a cutoff frequency
$\nu_{\rm therm}$ below which the spectrum will be thermalized:
\begin{equation}
\left(\frac{h\nu_{\rm therm}}{kT_{\rm shock}}\right) \approx
1.3\, \alpha_{-1}^{-23/20} \eta_{-1}^{-4/5}
\dot{m}_{-1}^{4/5} M_8^{-3/20} x_4^{-7/5} T_7^{-7/4}.
\end{equation}
If $\tau_{\rm therm} < 1$, the thermalized band will
always be in the Rayleigh-Jeans portion of the Planck spectrum.

The total energy content in this low-frequency thermalized portion of
the spectrum is given by
\begin{equation}
U_{\rm therm} \approx \frac{5}{\pi^4} \left(\frac{h\nu_{\rm
    therm}}{kT_{\rm shock}}\right)^3 aT_{\rm shock}^4.
\end{equation}
When this is comparable to the energy in the post-shock gas,
$3/2(\rho/\mu)k T_{\rm shock}$, the temperature of the matter must be
reduced by a factor of order unity in order to supply the energy in
the radiation, in turn raising $\tau_{\rm therm}$ ($\propto T^{-7/4}$)
and therefore $U_{\rm therm}$.  Further depression
of the gas temperature follows, leading to a still wider bandwidth of
thermalization.  Rapid approach to thermodynamic equilibrium is
thus the end-result whenever the initial energy in thermalized
photons is comparable to the gas energy.  This condition of
$U_{\rm therm} \gtrsim 3/2(\rho/\mu)k T_{\rm shock}$ will be satisfied for
all shock temperatures with
\begin{equation}
T_{\rm shock} \lesssim 2 \times 10^{10}\alpha_{-1}^{11/9}
\eta_{-1}^{-8/9} \dot{m}_{-1}^{8/9} M_8^{1/9} x_4^{-7/6}
\mbox{ K}.
\end{equation}
For our fiducial parameters, the corresponding shock speed is so
high---nearly $40,000$~km/s, roughly ten times the largest possible kick
speed---that highly unusual conditions are required for this condition
to be violated. Furthermore, since the shock speed
is limited by the orbital speed, and $V_{\rm orb}\approx
3000\, x_4^{-1/2}\mbox{ km/s}$, only systems with extremely low
accretion rates ($\dot{m}_{-1}^{8/9} \lesssim 10^{-3}$) have any
chance of avoiding rapid thermalization. But as we mentioned above, since the
luminosity is proportional to the disk surface density (and $\Sigma
\sim \dot{m}^{3/5}$), we expect to observe only those
systems with moderate-to-large $\dot{m}$ in the first place!

Even if free-free absorption were not enough to
thermalize the radiation, Compton scattering could also do the job
on a timescale roughly comparable to $t_{\rm rad}$.
This is because the characteristic Compton-$y$ parameter at
the post-shock temperature is
\begin{equation}
y = \frac{4 k T_{\rm shock}}{m_e c^2}\tau_{\rm es}^2
\approx 1 \times 10^6\, V_{1000}^2 \alpha_{-1}^{-8/5}
\eta_{-1}^{-6/5} \dot{m}_{-1}^{6/5} M_8^{2/5} x_4^{-6/5}.
\end{equation}
In this case, the thermal energy of the gas is drained as the
electrons scatter into a Wien spectrum the large number of low-energy
photons they have created by free-free emission.  The
timescale for this process is almost as fast as the free-free
radiation cooling time itself because each low-energy photon
receives on average an additional energy $\sim kT_e$ by subsequent
inverse Compton scattering.

\subsection{Light curves and spectra}\label{lcurves_spectra}

Thus, by any of several processes, the very large
column densities found in these disks ensure that thermalization is inevitable.
After the disk is shock-heated and reaches a radiation-dominated
thermal state,
it will expand adiabatically to a post-shock scale height
determined by hydrostatic equilibrium:
\begin{equation}
h^2_{\rm shock} \sim c_s^2\, T_{\rm orb}^2 \sim \frac{p}{\rho}
\frac{R^3}{GM},
\end{equation}
with the pressure and density given by $p=dE_{\rm kick}/(2\pi h_{\rm shock}\, R\,
dR)$ and $\rho=\Sigma/h_{\rm shock}$. Since the initial orbital energy of the 
annulus is $dE_0=(2\pi R\, dR\, \Sigma)(-GM/R)$, we can write the post-shock
scale height as
\begin{equation}
h_{\rm shock} \approx \left|\frac{dE_{\rm kick}}{dE_0}\right|^{1/2}R.
\end{equation}
The internal energy of the perturbed disk is then radiated over a cooling
time at each radius, which for optically thick,
radiation-pressure dominated disks is given by
\begin{equation}
t_{\rm cool} \approx \tau_{\rm es} \frac{h}{c}
\approx 0.33\, \Sigma(R)\, \left|\frac{dE_{\rm kick}}{dE_0}\right|^{1/2} \frac{R}{c}.
\end{equation}

The evolution of internal energy $dE_{\rm shock}$ due to intersecting
orbits of the perturbed disk is thus governed by
\begin{equation}
\frac{d}{dt}dE_{\rm shock}=\left. \frac{d E_{\rm kick}}{t_{\rm
    heat}}\right|_{t<t_{\rm heat}}-\frac{dE_{\rm shock}}{t_{\rm cool}},
\end{equation}
which gives a radiated luminosity of
\begin{equation}\label{dL_rt}
dL_{\rm shock}(R;t) = \frac{dE_{\rm kick}(R)}{t_{\rm heat}(R)} \times \left\{
\begin{array}{l@{\quad:\quad}l}
(1-e^{-t/t_{\rm cool}}) & t<t_{\rm heat} \\
(e^{t_{\rm heat}/t_{\rm cool}}-1)e^{-t/t_{\rm cool}} & t>t_{\rm heat}
\end{array} \right. .
\end{equation}
Thus, the luminosity grows linearly at early times, which is
reasonable for crossing eccentric orbits, and decays exponentially at
late times, appropriate for a diffusion process.

On top of this luminosity from the internal heating of the perturbed disk, we also
assume that outside of $R_{\rm in}$, the disk continues to behave as
in a steady-state accretion disk, removing energy and angular
momentum from the gas, giving the classical
accretion luminosity at each point in the disk:
\begin{equation}
dL_{\rm acc}(R) = \frac{3}{2}\frac{G M\dot{m}}{R^2}
\frac{L_{\rm Edd}}{\eta c^2} dR.
\end{equation}
While the shock-heating and subsequent expansion of the perturbed disk
may change the local accretion rate, we generally find the cooling
time to be much shorter than the inflow time, so we continue to assume
an overall steady-state disk model with constant $\dot{M}$ at all
radii. We anticipate that more detailed numerical simulations should
be able to model the dynamic behavior of these perturbed disks in the
near future.
The local emission spectrum is taken to be a thermal blackbody with
\begin{equation}\label{dL_tot}
dL_{\rm tot}(R)=dL_{\rm shock}+dL_{\rm acc}= 4\pi R\, dR\, \sigma T^4(R),
\end{equation}
with $\sigma$ the Stefan-Boltzmann constant. After the merger, the
inner edge of the disk continues to migrate inwards on an inflow
timescale, eventually closing the gap and forming a ``standard''
accretion flow in the inner regions of the disk.

\begin{figure}
\begin{center}
  \scalebox{1.0}{\includegraphics{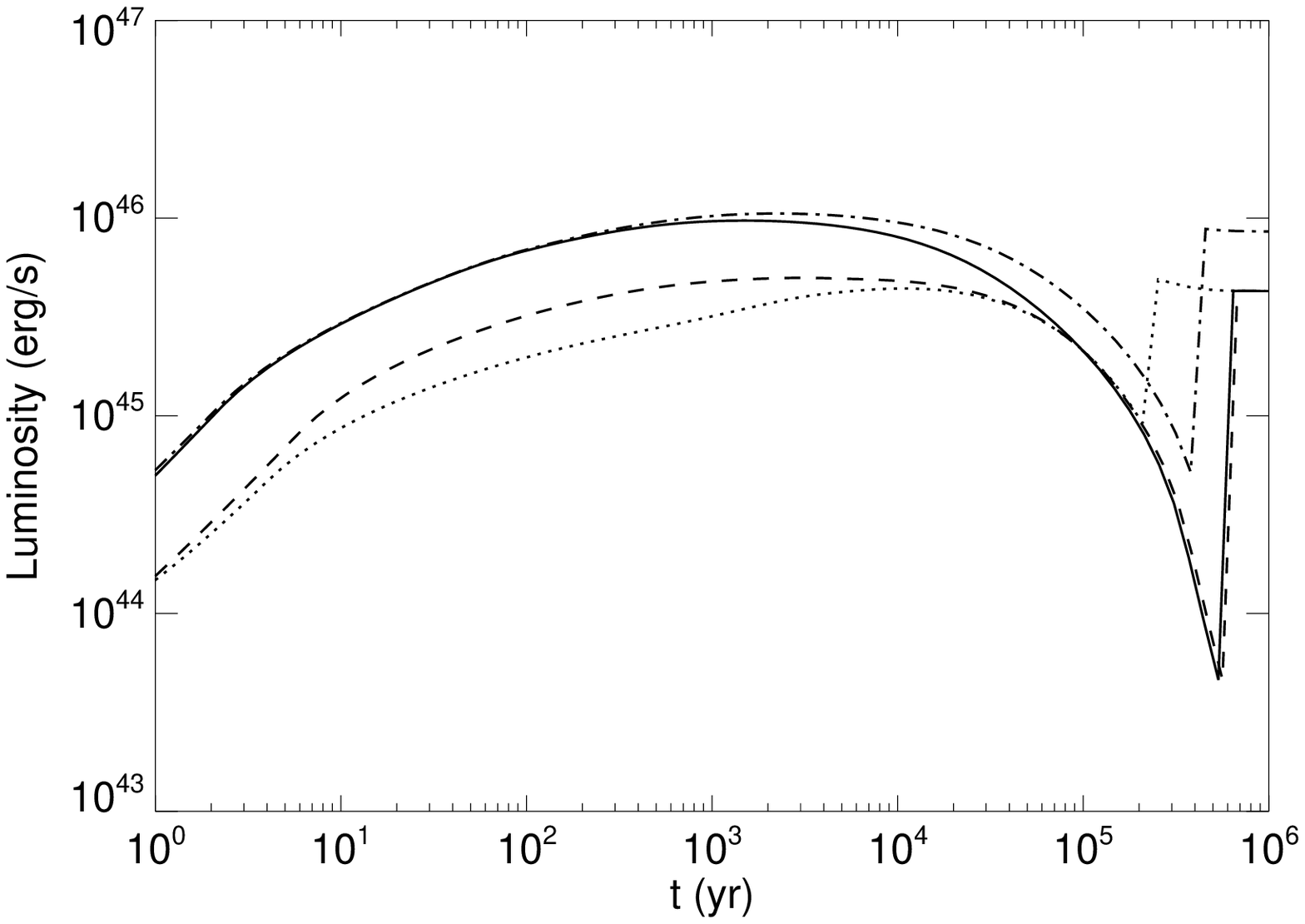}}
  \caption{\label{light_curves} Light curves from a few different
  characteristic mergers. In all cases, $M=10^8 M_\odot$, $\eta=0.1$,
  and $\alpha=0.1$. Solid line: $q=1$, $V_{\rm
  kick}=1000$ km/s, $\dot{m}=0.1$; Dashed line: $q=1$, $V_{\rm
  kick}=300$ km/s, $\dot{m}=0.1$; Dot-dashed line: $q=1$, $V_{\rm
  kick}=1000$ km/s, $\dot{m}=0.2$; Dotted line: $q=0.1$, $V_{\rm
  kick}=300$ km/s, $\dot{m}=0.1$. The step at late times is due to the
  gap in the disk filling in (this happens earlier for smaller mass
  ratios and higher accretion rates).}
\end{center}
\end{figure}

Figure \ref{light_curves} shows
light curves for several events with $M=10^8 M_{\odot}$,
$\eta=0.1$, and $\alpha=0.1$, but varying $q$, $V_{\rm kick}$, and
$\dot{m}$. In all cases, we see an
initial linear rise in luminosity as the gas near $R_{\rm in}$ is
first excited and then radiates its heat on a timescale of a few
years.
The peak luminosity is typically reached after $\sim 10^3$~yr,
and the afterglow remains roughly this bright for $\sim 10^4$~yr.
In the solid, dashed, and dot-dashed
curves, the binary mass ratio is unity, giving an inner edge of
$R_{\rm in} \sim 10^3 M$. In these cases, the inner gap is not closed
until nearly $10^6$ years after merger. At that point, the luminosity
increases sharply. The reason for this abrupt behavior is that the
inner edge of the disk moves in very slowly at large $R$, then
accelerates its inward motion as the gap closes, leading to a nearly
instantaneous increase in total luminosity.
The solid and dashed curves have
$\dot{m}=0.1$, while the dot-dashed curve has $\dot{m}=0.2$. For a
higher accretion rate, $R_{\rm in}$ moves in slightly, and from
equation (\ref{Sig_r}), we see that overall mass--and thus
luminosity--of the perturbed disk also increase with $\dot{m}$.

The solid and dot-dashed curves in Figure \ref{light_curves} correspond to a recoil
of $V_{\rm kick}=1000$ km/s,
and the dashed and dotted curves have $V_{\rm kick}=300$ km/s.
From equation (\ref{dE}) we see that for the inner regions of
the disk, where $GM\, \epsilon_{\rm GW}/R \gg V_{\rm kick}^2$, the
perturbations due to
mass loss are greater than those due to the kick, while in the outer
regions, the kick term dominates.
The dotted curve has $q=0.1$, which causes the inner edge of
the disk to be closer in at the time of decoupling, so the delay
before gap filling is also shorter. From equation (\ref{e_GW}), we
expect a smaller GW mass loss, and so the initial heating rate is
smaller. As can be seen from equation (\ref{dL_rt}), the inner disk lights up on
roughly the same timescale as the $q=1$ cases despite having a smaller
value of $R_{\rm in}$. Note that all cases
eventually settle down to nearly the same luminosity after the gap
closes because the luminosity in the normal AGN phase is determined
only by $\dot{m}$ and $M$.

\begin{figure}
\begin{center}
  \scalebox{1.0}{\includegraphics{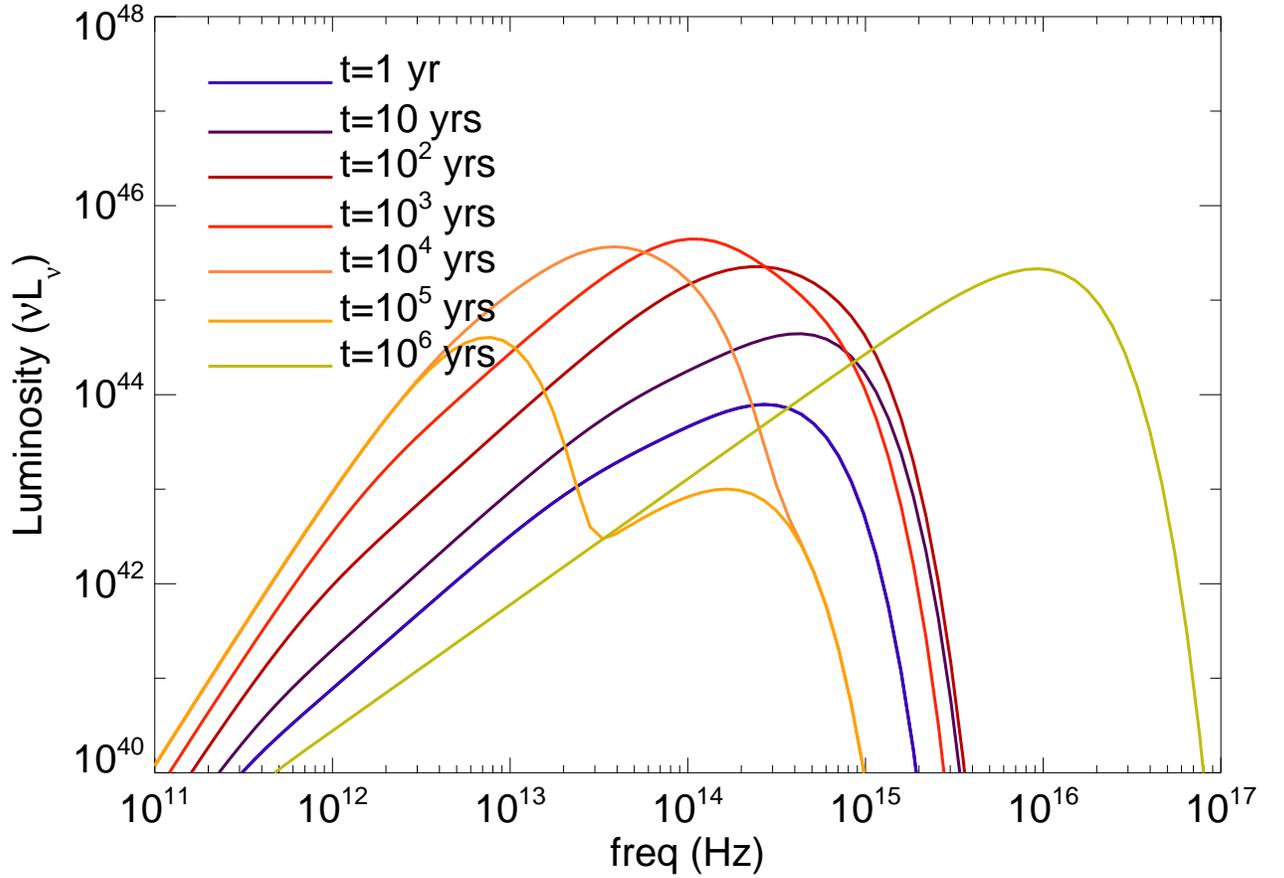}}
  \caption{\label{spec_time} Spectra of perturbed disk at a series of
  different times for $M=10^8 M_\odot$, $\dot{m}=0.1$,
  $q=1$, and $V_{\rm kick}=1000$ km/s. The disk brightens at early
  time, then reddens while dimming slightly before brightening again
  as the inner disk lights up.}
\end{center}
\end{figure}

In Figure \ref{spec_time} we show a time sequence of spectra from a merger with
$M=10^8 M_\odot$, $\dot{m}=0.1$, $q=1$, and $V_{\rm
  kick}=1000$ km/s (the solid curve in Fig.\ \ref{light_curves}). Over
the first hundred years or so, the disk brightens as
the perturbed orbits near the inner edge begin to intersect and dissipate
energy. Then, as the region of maximum dissipation propagates outward,
the disk slowly dims and reddens, its radiation moving from the optical/UV to IR
over thousands of years. As in Figure \ref{light_curves}, at around $10^6$
years after the
merger, the inner gap closes, forming a ``typical'' AGN with thermal
emission peaking in the UV band.

At even later times, the situation becomes highly uncertain, as the
system's behavior depends
largely on the existence and state of gas in the
outermost regions of the disk ($R \gtrsim 10^5 M$). Even within this
radius, it is not clear whether the gas can be described by a simple
thin disk model as in \citet{shakura73}. Beyond $R\gtrsim 10^4 M$, the
disk may become gravitationally unstable, but there appears to be
significant observational evidence that complete fragmentation and
collapse is avoided by some auxiliary heating mechanism [\citet{lodato08}
and references therein]. In our simple model, if there is in fact
appreciable gas and
thus emission beyond $R\sim 10^5 M$, it should appear as a steadily
reddening peak in the far infrared, as might be extrapolated from the
curves in Figure \ref{spec_time}. However, even if there is a
significant amount of gas in these outermost regions, it will almost
certainly be unbound after any appreciable kick (orbital speeds at
$R\sim 10^6 M$ are a few hundred km/s), and thus inherently limited in
its total available energy content.

To estimate the light curves and spectra we might expect to see from
LISA counterparts with $M\sim 10^6 M_\odot$, recall that the
total energy released in the afterglow scales like $\Sigma R^2 \sim
M^{11/5}$ and the typical timescales vary as $M$. Thus the peak
luminosity should be a factor of $\sim 10^{-2}$ smaller than those
plotted in Figure \ref{light_curves} and evolve on a timescale 100
times shorter. The surface brightness, however, which scales like
$\Sigma$, has only weak dependence on the total mass, so we expect
LISA afterglows to have similar spectra, peaking in the near- to
mid-IR over a period of a few years after the merger.

\section{COSMOLOGICAL MERGER HISTORY}
\label{cosmological_model}

\begin{figure}
\begin{center}
  \scalebox{0.8}{\includegraphics{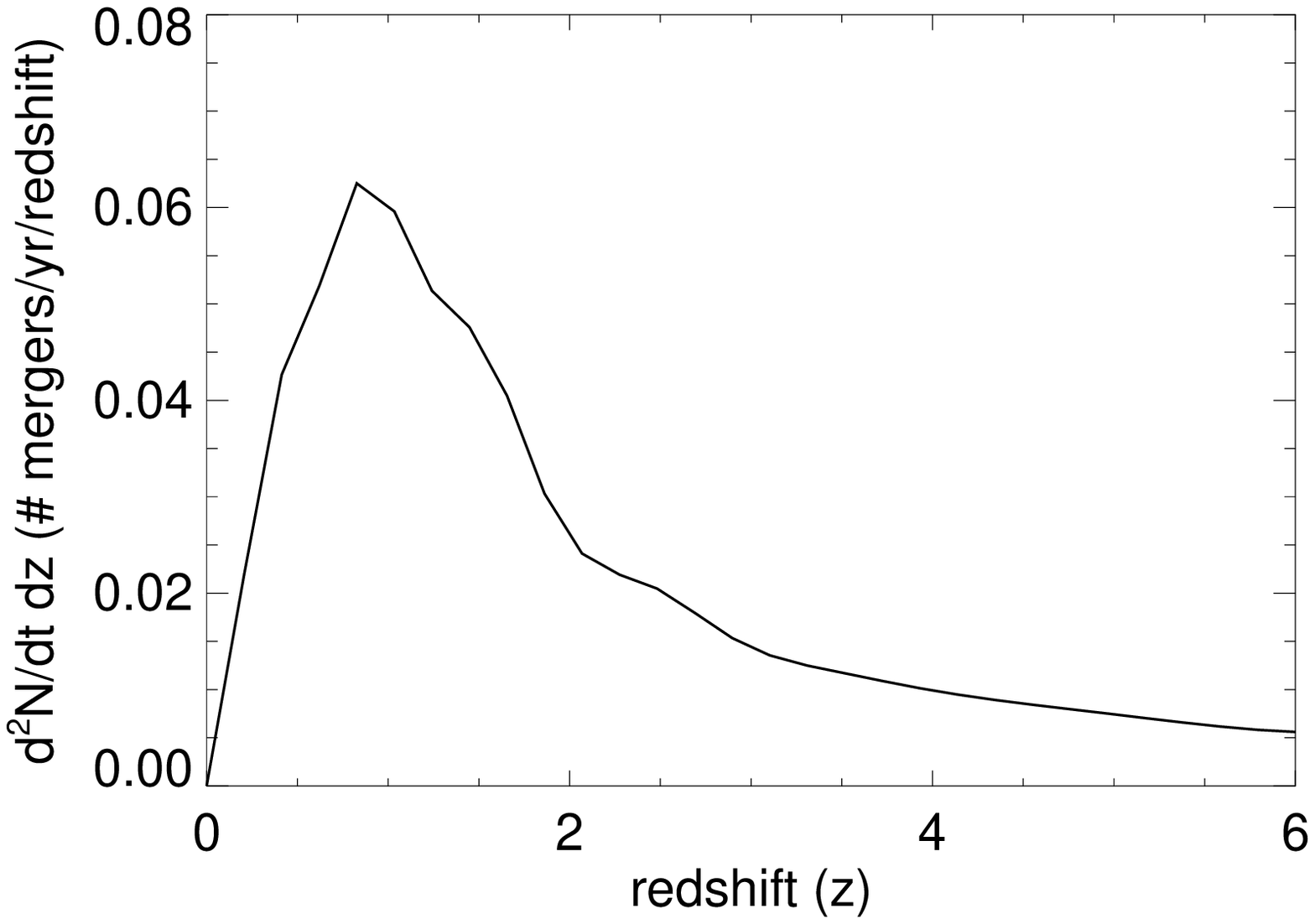}}
  \scalebox{0.8}{\includegraphics{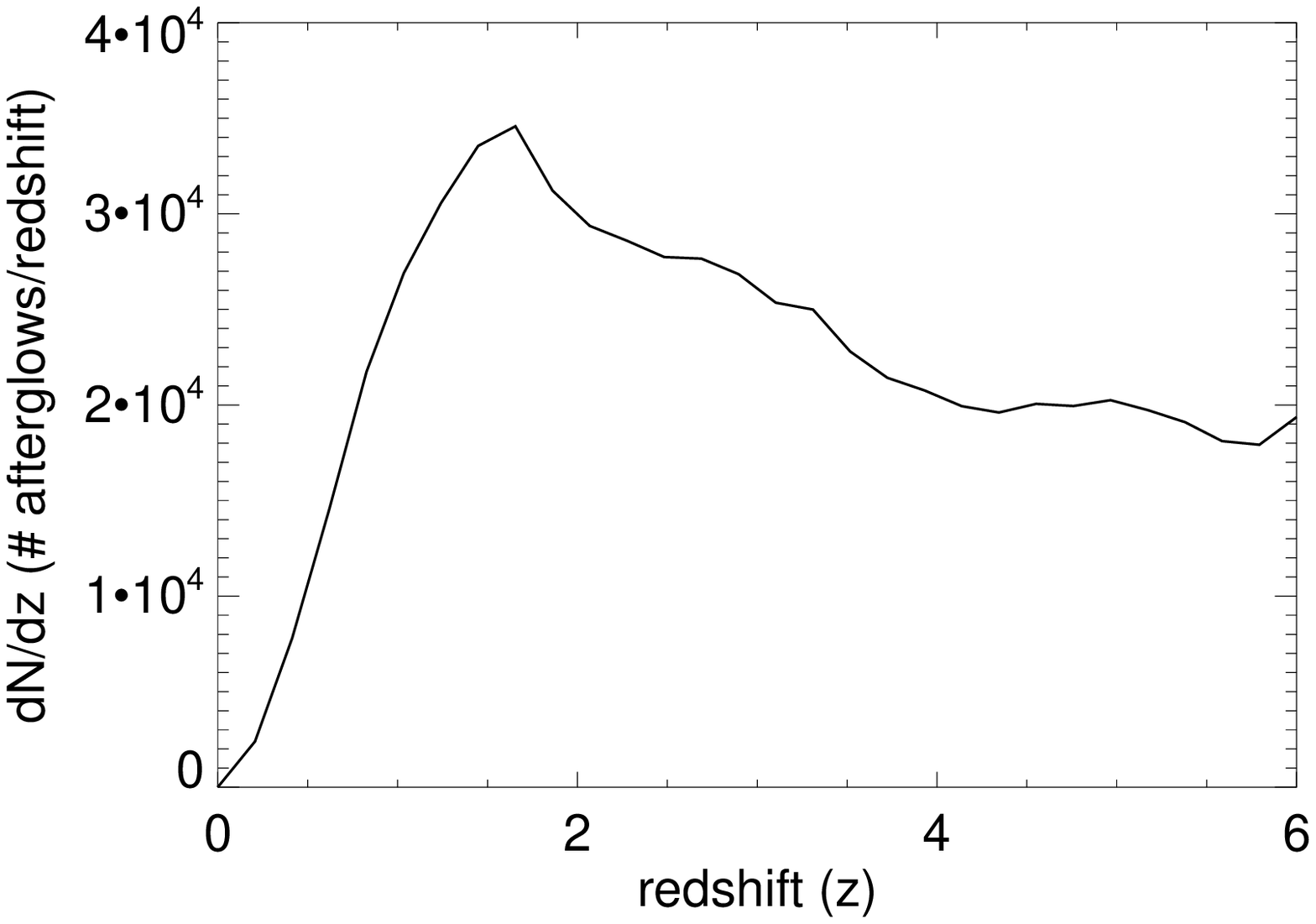}}
  \caption{\label{Nmerge} (top) Merger rate of SMBHs in observable
  universe per year per unit redshift, assuming each BH undergoes an
  average of three major mergers between $z=6$ and $z=0$. (bottom) Total number of
  potentially observable afterglows per unit redshift. An afterglowing
  system is one whose luminosity is dominated by the thermal relaxation
  of the perturbed disk and where the central gap has not yet closed.}
\end{center}
\end{figure}

We adopt a standard $\Lambda$CDM cosmology with $\Omega_\Lambda=0.74$,
$\Omega_M=0.26$, and normalized Hubble constant $h=0.73$
\citep{spergel07}. At each
redshift, we approximate the BH mass function by the anti-hierarchical
distributions given by \citet{merloni04}, extrapolated to redshift
$z=6$. We assume that each BH experiences an
average of $N_{\rm mm}$ major mergers between $z=6$ and the present
and that over this range, the merger rate per comoving volume is
weakly dependent on redshift \citep{sesana04}.
The top panel of Figure \ref{Nmerge} shows the
expected number of mergers with $M>10^6M_\odot$ throughout the
universe per observer year per unit redshift for $N_{\rm mm}=3$
(in rough agreement with the lower-right panel of
  Fig.~1 in \citet{sesana04}). All rates quoted
below should simply scale linearly with this parameter $N_{\rm mm}$. Note that
these merger rates are significantly lower than those typically quoted
for LISA sources, which include a large number of smaller
black holes ($M\lesssim 10^6 M_\odot$) at high redshift ($z\gtrsim
10$) \citep{menou01,sesana04,rhook05}. 

While the
  mass functions given by \citet{merloni04} likely underestimate the
  number of low-mass AGNs due to observational selection effects,
  these systems are also much less likely to produce long-lived, bright
  afterglow signals. Therefore, in estimating number counts, it is
  reasonable to use the observations of \citet{merloni04}
  to normalize the high-mass portion of the distribution function,
  even though they may be interpreted in terms of anti-hierarchical
  evolution, and
  use the hierarchical simulations of \citet{sesana04} to estimate the
  merger rates of these high-mass systems as a function of
  redshift. As mentioned above in Section \ref{lcurves_spectra}, even if the number of low-mass ($M\lesssim 10^6
  M_\odot$) binaries is greater by an order of magnitude, their
  luminosities and lifetimes will also be much smaller than those
  systems with mass around $10^8 M_\odot$, so the observable IR luminosity
  function should be rather insensitive to mergers involving low-mass
  black holes.

For a given merger, we determine the masses by selecting two black
holes randomly from the distribution $\Phi(M,z)$ given in
\citet{merloni04}. The kick magnitude
is determined from the formulae given by \citet{baker08}, which agree
with earlier estimates of \citet{schnittman07} for small to moderate
kicks, but predict a larger number of extreme kicks above 1000 km/s. We assume a
uniform distribution of BH spin orientations and set $a/M=0.9$ for all
systems. Recoil velocities generally scale
linearly with the spin parameter, so should not be too sensitive to
small uncertainties in this typical value. With regard to the spin
orientation, \citet{bogdanovic07} recently argued that BHs embedded in a
circumbinary accretion disk should rapidly align their spins with the
overall angular momentum of the disk, which would in turn lead to much
smaller recoil values. On the other hand, if the disk forms outside of
the binary, we expect very little gas to actually accrete directly
onto either black hole, and the BHs
could very well maintain their original random orientations
\citep{schnittman04}.

While the vast majority of isolated SMBHs are not active
at any given time, the accretion rates for merging BHs may on average be
much higher due to the inflow of gas after galactic
mergers \citep{haehnelt93,mihos94,volonteri03}. Furthermore, many models rely on the
dynamical friction of gas disks to bring the
BHs close enough for gravitational radiation to take over and merge
within a Hubble time---the ``final parsec problem'' \citep{escala05,dotti07}. Thus, we
assume every merging system has a significant circumbinary disk,
with accretion rates uniformly distributed in logarithmic space
between $\dot{m}=0.01$ and $1$.

For each merger, we define the afterglow
phase as the time during which the total luminosity in equation
(\ref{dL_tot}) is dominated by the shock heating due to the perturbed
orbits and where the inner gap has not yet closed.
In the bottom panel of Figure \ref{Nmerge} we show
the expected number of objects in the
afterglow phase observable at the present time as a function of redshift. Note
that while the merger rate (top panel of Fig.\ \ref{Nmerge})
decreases with $z\gtrsim 1$, the redshifted time makes those events
appear to last longer to us, and thus there should be more in the
afterglow phase at any given time. Additionally, due to the
anti-hierarchical growth of SMBHs, mergers at higher
redshifts typically involve larger masses and longer afterglows.

\begin{figure}
\begin{center}
  \scalebox{1.0}{\includegraphics{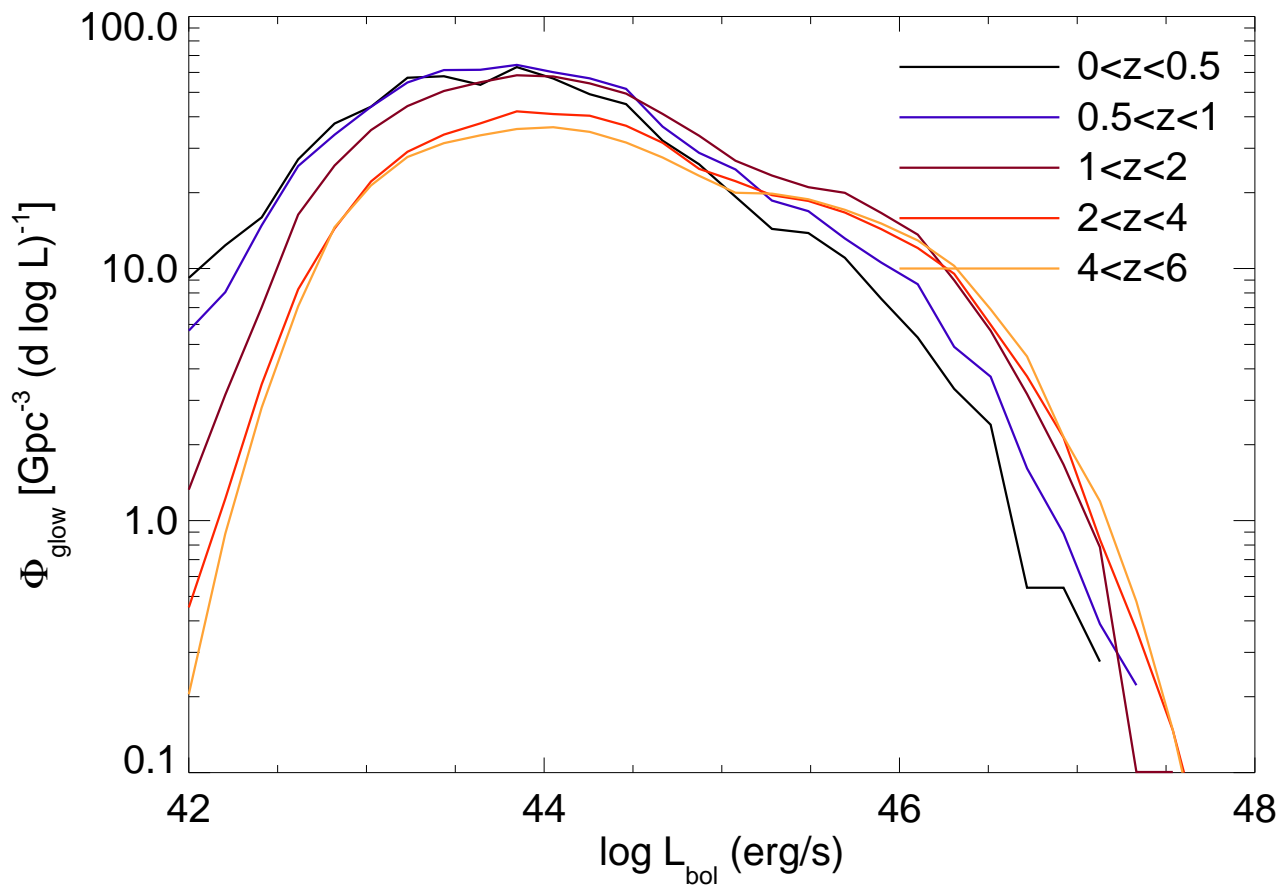}}
  \caption{\label{Phi_L} Luminosity distribution function (number per
    comoving volume per log luminosity) for systems
  in the afterglow phase at different redshifts. As the number of
  low-mass BHs ($M \lesssim 10^7 M_\odot$) increases at smaller redshifts,
  the distribution function shifts to slightly lower luminosities.}
\end{center}
\end{figure}

In Figure \ref{Phi_L} we show the luminosity function $\Phi_{\rm
glow}(L,z)$ (in units of number per comoving volume per log
luminosity) for systems in
the afterglow phase in a number of different redshift bins. The
distribution is remarkably narrow and constant in time. There are a
number of reasons for this behavior. First, the BH mass distribution
we use is peaked around $10^8 M_\odot$ and falls off sharply above
$10^9 M_\odot$ \citep{merloni04}. In the downsizing paradigm of cosmic
evolution, the
number of smaller-mass BHs ($\lesssim 10^7 M_\odot$) increases at low
redshift, in turn giving a slightly larger number of low-luminosity afterglows,
as shown by the black and blue curves in Figure
\ref{Phi_L}. However, these low-mass systems are also much
shorter-lived in their afterglow phase, giving a strong selection
effect against seeing them at any one time. The mass accretion rate
$\dot{m}$ is also limited by a selection effect:
we consider only $\dot{m}\ge 0.01$ in our model because if $\dot{m}$
is too low (thus giving a low-luminosity, low-mass disk), the BH binary
does not evolve quickly enough via dynamical friction with the disk to
merge within a Hubble time \citep{escala05,dotti07}. If, however,
$\dot{m}$ is too large (high-luminosity), the inflow time at the inner
edge of the disk [eqns. (\ref{t_inflow2}, \ref{t_inflow3})] will be
small enough such that no appreciable gap is formed and the afterglow
is not readily identifiable. Additionally, we see from equations
(\ref{dE}) and (\ref{Sig_r}) that the total energy released is a relatively
weak function of $\dot{m}$, further narrowing the range of
luminosities. Lastly, the mass ratio $q$ is
likely to be close to unity for a similar reason: if it is too small,
the GW inspiral time is longer than the gas inflow time and the gap is
small and short-lived. This selection for large $q$ also favors higher
kick velocities, which for spinning BHs are maximized when $q=1$
\citep{schnittman07, baker08}.

\section{OBSERVATIONAL POTENTIAL}
\label{observations}

\begin{figure}
\begin{center}
  \scalebox{1.0}{\includegraphics{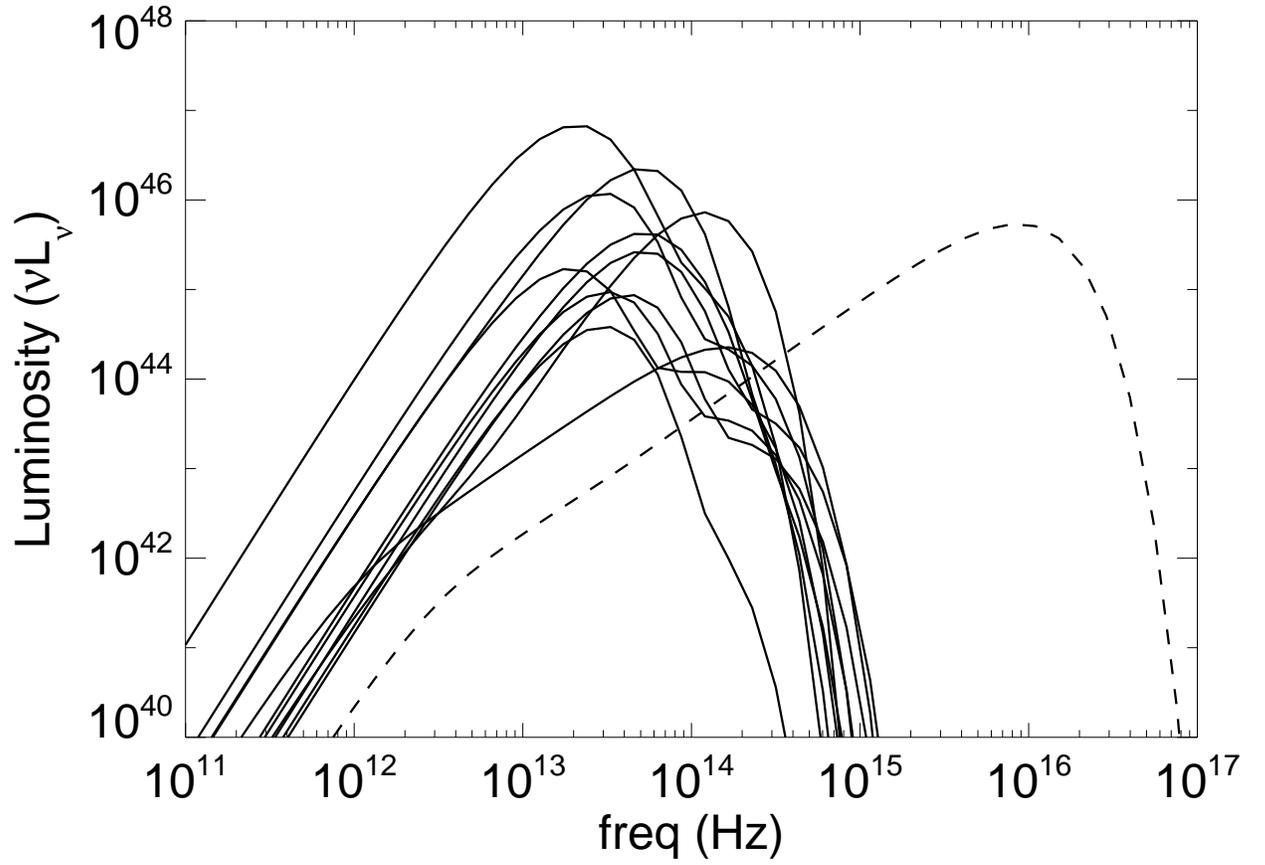}}
  \caption{\label{Fnu_all} Rest-frame spectra of a sample of afterglow
  systems at any one time (solid curves). For reference, the dashed
  curve is the spectrum of a similar merger system shortly after the
  central gap has closed.}
\end{center}
\end{figure}

Because any single merger event spends the vast majority of its
lifetime in the late, IR-dominated phase corresponding to the outer
disk relaxation, the average system seen today should be
characterized by the double-hump spectra shown as the orange and
yellow curves in Figure \ref{spec_time}. Using the cosmological model
parameters from the previous
section, in Figure \ref{Fnu_all} we show a randomly selected sample of
rest-frame spectra within
$\sim 10^6$ years after merger. The solid curves correspond to those
systems formally in the afterglow phase, when the luminosity is
dominated by the dissipation from the kicked disk. For reference, the dashed curve
is from a system that has already closed its central gap and
radiates as a normal AGN in the inner regions. We find that, for the
model parameters used above, the numbers of closed-gap and open-gap
systems are roughly equal, with somewhat more closed-gap systems at
low redshift, where the typical BH masses---and thus timescales---are
smaller.

Closed-gap systems are essentially identical to ``normal'' AGN,
dominated by a thermal peak in the UV from the inner disk, along with
a strong IR peak due to reprocessing
of the UV by a surrounding dusty torus at large distance, which could
easily be confused with the IR afterglow emission from the perturbed
outer disk. Thus, for
identification purposes, we will focus on kicked disks where the
central gap has not yet closed, so the emission is almost entirely in the
IR. Due to the lack of a central engine, these systems are not likely
to produce significant UV/X-ray flux or compact radio jets. Furthermore, we
expect their time variability to be quite low, since most of the
emission is coming from the outer regions of the disk, where the
dynamical time is  hundreds of years.

\begin{figure}
\begin{center}
  \scalebox{1}{\includegraphics{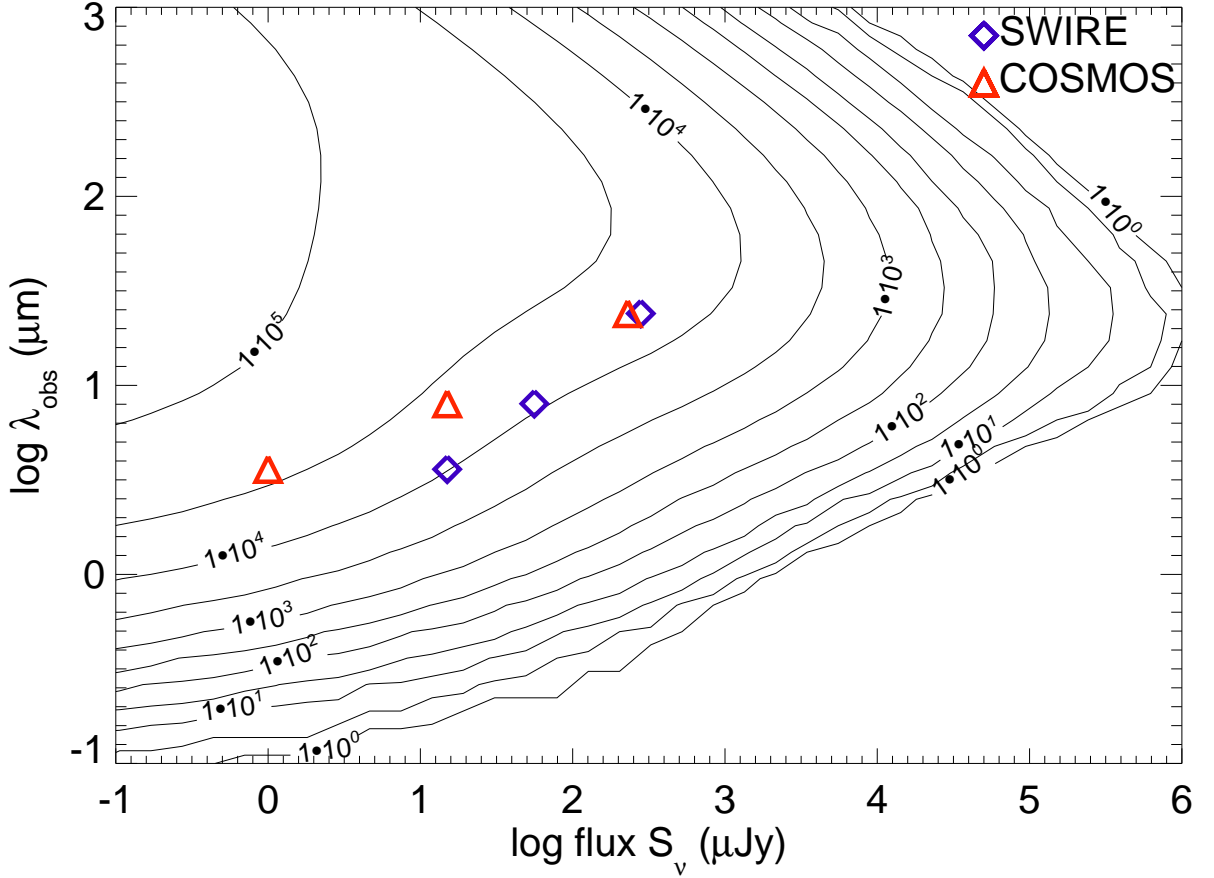}}
  \caption{\label{contN_Snu} Contour plot of source counts $N(>S)$
  in the entire sky as a function of observed wavelength (vertical
  axis) and flux (horizontal axis). Also shown are the flux limits for
  the SWIRE (blue diamonds) and COSMOS (red triangles) surveys. More
  massive systems tend to be
  brighter and last longer, so the $N(>S)$ relation is flatter than the
  standard $S^{-3/2}$ power-law for uniform source distributions.}
\end{center}
\end{figure}

Restricting
ourselves to this sample of open-gap disks, in Figure
\ref{contN_Snu} we show the source counts in the sky over a range of
observed wavelengths, plotting contours of $N(>S_\nu,\lambda_{\rm obs})$.
Also shown in Figure \ref{contN_Snu} are the $5\sigma$ flux limits from the
wide-field, multi-band
SWIRE and COSMOS surveys.
SWIRE has a total coverage of nearly 50 deg$^2$, so
on the basis of Figure \ref{contN_Snu} we might expect to
find $\sim 10$ afterglowing sources in the SWIRE field.
The COSMOS survey covers a somewhat smaller area, 2 deg$^2$,
with sensitivity comparable to SWIRE at 24 $\mu$m, but significantly better
at 3.6 and 8 $\mu$m, so there may again be $\sim 1-10$ afterglow sources
detectable in its field.
The GOODS survey goes considerably fainter, to
40 $\mu{\rm Jy}$ at 24 $\mu$m, but covers an area of just $150$ arcmin$^2$,
giving an expected number of sources of only $\sim 0.01-0.1$.

In order to distinguish these afterglow signals from other
ultra-luminous IR sources, we require very low flux limits from
optical, UV, and X-ray observations. As an initial estimate, we might
require an X-ray flux no greater than 10\% of the total IR flux,
corresponding to roughly $10^{-14}-10^{-13}$ ergs/cm$^2$/sec for the
typical afterglowing source within $z\lesssim 3$. This is well within the
limits of the {\it XMM-Newton} Large Scale Structure and the {\it Chandra}
XBootes surveys, which together cover roughly 18 deg$^2$ of the
SWIRE field. COSMOS has somewhat deeper coverage with {\it XMM}, as well as full
{\it HST} coverage with ACS, so may be the preferable approach at this point.

\begin{figure}
\begin{center}
  \scalebox{0.5}{\includegraphics{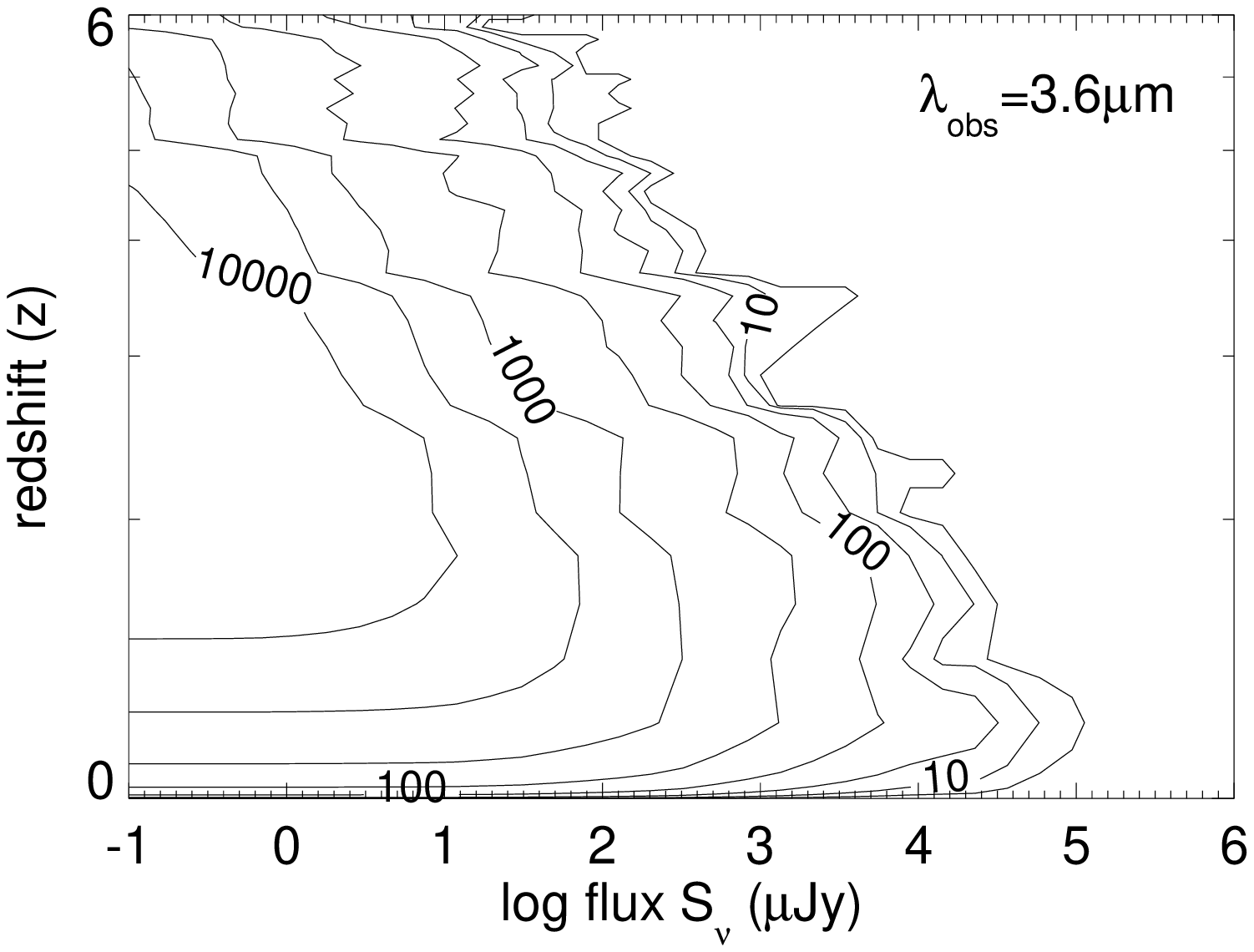}}\\
  \scalebox{0.5}{\includegraphics{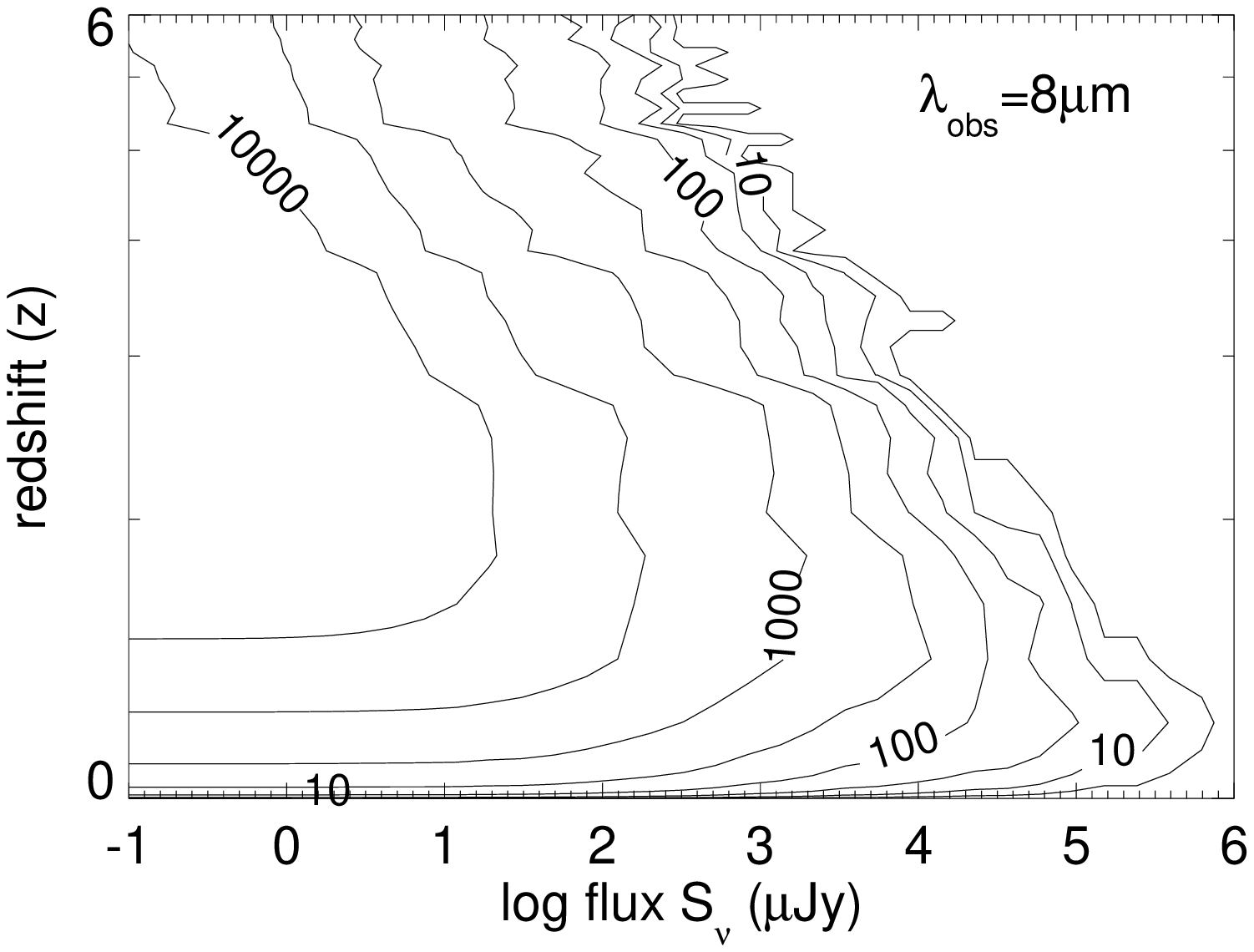}}\\
  \scalebox{0.5}{\includegraphics{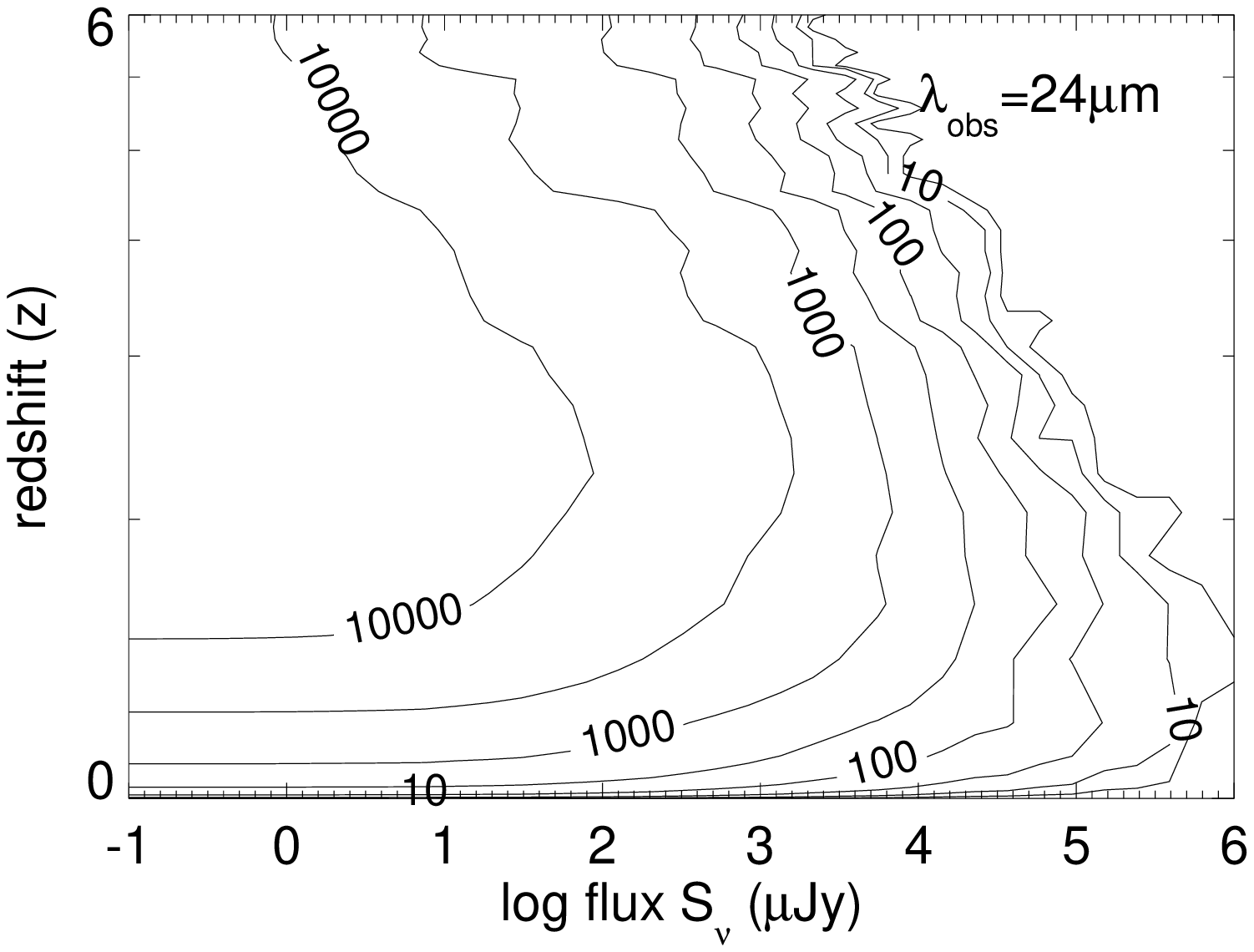}}
  \caption{\label{contN_Sz} Source distribution contour plot as a
  function of redshift and IR flux, observed at (top to bottom) 3.6, 8,
  and 24$\mu m$. The contours are measures of $N(>S)$ per unit redshift.}
\end{center}
\end{figure}

As we can
see from Figure \ref{contN_Snu}, in the mid-IR, the $N(>S)$ curves scale roughly as
$S^{-1}$, giving a larger number of high-flux sources than a uniform
distribution with no cosmic evolution ($N\sim S^{-3/2}$). Thus the
number of potentially
confusing sources should decrease with increasing flux, yet detection will
require covering an observed solid angle that increases linearly with
flux. In light of these trade-offs, we believe the best discrimination potential may
ultimately come from an extremely wide-field survey, even if shallow. To
estimate the distance and brightness of the high-flux end of the
distribution, in Figure
\ref{contN_Sz} we show source-count contours as a function of
redshift, finding that the majority of bright sources are indeed
nearby, with $z\lesssim 1$.

Many IR/X-ray selected candidates are likely to be obscured
AGN with a high Compton depth, which may be ruled out as
afterglow sources by the detection of hard X-rays or compact radio jets.
In addition to demonstrating a low level of X-ray flux,
we may need optical
spectroscopy to rule out the existence of line emission from UV
excitation of the region around the AGN. This may require targeted
spectroscopic observations, focusing on candidates selected with the
other criteria described above.

\section{DISCUSSION}
\label{discussion}

We have proposed the existence of a new class of electromagnetic
sources produced by the merger of two supermassive black holes and
thus corresponding to some of the strongest gravitational wave signals
in the observable universe. These objects are potentially observable today, long
before any planned GW experiment might detect them. When surrounded by a circumbinary
accretion disk, the two BHs can be driven towards merger via dynamical
friction. Eventually the gravitational wave losses dominate
the evolution of the binary, shrinking the orbit faster than the gas
inflow time, at which point the disk decouples from the binary,
leaving an open gap of extremely low-density gas and negligible direct
accretion onto the BHs. This decoupling is followed soon after by the
merger of the two BHs, disrupting the gas disk by the energy and
momentum losses in the gravitational waves. The perturbed, eccentric
orbits then intersect, forming shocks and heating the disk, which
then radiates its internal energy over a cooling time. The large
optical depth in the disk ensures that however the heat is released
initially, it will rapidly reach thermal equilibrium and should therefore
emit primarily in the infrared over most of the
afterglow's lifetime. Furthermore, {\it only} systems with high optical
depth will also have sufficient mass
in the disk to produce significant luminosities from internal shocks.
Due in part to the large optical depths of the disks, these IR
afterglow signatures could last hundreds of thousands of years for typical
BH masses of $M\sim 10^7-10^9 M_\odot$. At some point long after the
merger, the inner edge of the disk will migrate in towards the central
BH, forming a normal AGN and marking the end of the afterglow
phase.

Folding this perturbed disk model with a cosmological merger
scenario, we predict the expected event rate ($\sim 0.1$/yr) and total
number counts ($\sim 10^5$) for afterglowing disks observable in the
universe today.
Considering only existing survey data from {\it Spitzer}, {\it XMM},
and {\it Chandra},
roughly $1-10$ open-gap afterglow systems may be detectable
today. However, these sources would need to be discriminated from a
much larger population of deeply obscured AGN, either
with hard X-ray observations or
optical spectra that might rule out the existence of narrow-line
regions characteristic of normal AGN. With a large fraction of their
luminosity coming from the outer regions of the disk, where the
dynamical time is hundreds of years, we also expect
to see relatively low levels of variability in their light curves.

In addition to the observational challenges in successfully detecting
and identifying
these systems, there are also a number of theoretical
uncertainties in the underlying afterglow model. Most critically, the
properties of the gas disk (if it even exists!) at large radii will
strongly affect the predicted light curves and spectra described in
Section \ref{lcurves_spectra}. At this point, we have relatively little
understanding of AGN disks beyond $R\sim 10^4 M$, where much of the
afterglow emission originates. Even if the disk reliably extends out
to $R\gtrsim 10^5 M$, the gas accretion rate as well as the
surface density and temperature in these regions may very well not be
described by a simple $\alpha$-disk model.

Although we have used a nominal
accretion efficiency of $\eta=0.1$ and a stress coefficient
$\alpha=0.1$ throughout this paper, these
parameters are also uncertain, although by no more than a factor of a
few. They influence our predictions in only two ways: the total energy
deposited in the disk and also the radiative cooling time are
proportional to $\Sigma \propto \alpha^{-4/5} \eta^{-3/5}$, so the
total luminosity is independent of either parameter.
The inflow time---and thus the lifetime of the
afterglow phase---is proportional to $\alpha^{-1}$, so uncertainty in
$\alpha$ of order unity could vary the total number of observable
sources by a similar factor. As discussed
above at the end of Section \ref{cosmological_model}, the uncertainty
in $\dot{m}$ may not be too great, since strong selection effects limit
the likelihood of observing either very small or very large
$\dot{m}$. However, if the loss cone of scattering stars in the
galactic center can
be continuously replenished, it may be possible to shrink the BH binary
sufficiently through three-body interactions alone to trigger a GW
merger without a circumbinary disk.
If so, there may be additional black hole mergers without
enough surrounding gas to be as bright as the ones described here.
At the other extreme, there may
always be such a large amount of gas driving the merger that no
appreciable gap is ever formed in the disk, making the afterglow very
difficult to distinguish from a normal AGN.


The simple model we use for merger rates is also uncertain by at least a factor
of a few, but this should only affect the overall number counts of
sources, and not directly change their appearance. However, if the
BH masses in any given merger are not really selected at random from
the overall mass distribution function (e.g.\ comparable mass mergers
are favored due to the shorter time from galaxy merger to BH
merger), the number of afterglows will increase, since systems with
large $q$ tend to fill in their gaps more slowly. Finally, the
distribution of BH spin orientations is still quite uncertain, and
there may be significant evolution effects that favor aligned spins
with relatively small kicks, in turn reducing the average luminosity of the
afterglow population.

Of course, wherever there are theoretical uncertainties, there are
great observational opportunities. If numerous IR afterglows are
successfully detected and positively identified as open-gap accretion
disks, we can begin to use them as observational tools to probe SMBH
binary systems. Some of the important questions that may be answered
are: what is the distribution of BH spins and orientations? what is
the range of astrophysically relevant kick velocities? what are the
merger rates, total masses, and expected mass
ratios for SMBH binaries at various redshifts? how far out do
accretion disks extend and what are their properties at large
radii?

\vspace{0.35cm}\noindent We would like to thank Tamara Bogdanovic,
Cole Miller, Tim Heckman, and Marta Volonteri for helpful discussions and
comments. We would also like to thank the anonymous referee for their
close reading of the text and very constructive comments. This work
was supported by the Chandra Postdoctoral Fellowship Program (JDS) and
NSF grant AST-0507455 (JHK).


\clearpage

\end{document}